\documentclass[12pt]{article}
\usepackage{amsmath,amsfonts,amssymb,amsthm}
\usepackage{xcolor}
\usepackage{array, multirow,float}
\usepackage{pifont}

\usepackage{graphicx, subfig}
\usepackage{enumerate}
\usepackage{natbib, url, bm}
\global\long\def\bbeta{\boldsymbol{\beta}}%
\global\long\def\sstar{s^{*}}%

\newtheorem{theorem}{Theorem}[section]

\newtheorem{remark}{Remark}[section]
\DeclareMathOperator*{\argminB}{argmin}

\newcommand{\ve}[1]{{\mbox{\boldmath ${#1}$}}}

\newcommand{\blind}{1}

\begin{document}

\def\spacingset#1{\renewcommand{\baselinestretch}
{#1}\small\normalsize} \spacingset{1}


\if1\blind
{
  \title{\bf CoMMiT: Co-informed inference of microbiome-metabolome interactions via transfer learning}
  \author{Leiyue Li, 
    \hspace{.2cm}\\
    Department of Math, Computer Science, and Statistics, St. Lawrence University\\
    Chenglong Ye, \\
    Department of Statistics, University of Kentucky\\
    Tim Randolph, \\
    Clinical Research Division, Fred Hutch Cancer Center\\
    Meredith Hullar, Johanna Lampe, Marian Neuhouser\\
    Public Health Sciences Division, Fred Hutch Cancer Center\\
    Daniel Raftery, \\
    School of Medicine, University of Washington\\
    and \\
    Yue Wang, \\
    Department of Biostatistics and Informatics, University of Colorado
    }
  \maketitle
} \fi

\if0\blind
{
  \bigskip
  \bigskip
  \bigskip
  \begin{center}
    {\large \bf CoMMiT: Co-informed inference of microbiome-metabolome interactions via transfer learning}
\end{center}
  \medskip
} \fi

\bigskip
\begin{abstract}

Recent multi-omic microbiome studies enable integrative analysis of microbes and metabolites, uncovering their associations with various host conditions. Such analyses require multivariate models capable of accounting for the complex correlation structures between microbes and metabolites. 
However, existing multivariate models often suffer from low statistical power for detecting microbiome-metabolome interactions due to small sample sizes and weak biological signals.
To address these challenges, we introduce CoMMiT, Co-informed inference of Microbiome-Metabolome Interactions via novel Transfer learning models. 
Unlike conventional transfer-learning methods that borrow information from external datasets, CoMMiT leverages similarities across metabolites within a single cohort, reducing the risk of negative transfer often caused by differences in sequencing platforms and bioinformatic pipelines across studies.
CoMMiT operates under the flexible assumption that auxiliary metabolites are collectively informative for the target metabolite, without requiring individual auxiliary metabolites to be informative. 
CoMMiT uses a novel data-driven approach to selecting the optimal set of auxiliary metabolites. 
Using this optimal set, CoMMiT employs a de-biasing framework to enable efficient calculation of p-values, facilitating the identification of statistically significant microbiome-metabolome interactions.
Applying CoMMiT to a feeding study reveals biologically meaningful microbiome-metabolome interactions under a low glycemic load diet, demonstrating the diet-host link through gut metabolism.
 
\end{abstract}

\noindent%
\vfill

\newpage
\spacingset{1.9} 
\section{Introduction}
\label{sec:intro}

The Human Microbiome Project (HMP) conducted a comprehensive study of the healthy human microbiome across multiple body sites to identify a common microbial profile for healthy individuals, serving as a benchmark for diseased populations \citep{human2012structure_new}. Among these sites, the gut microbiome has received significant attention due to its strong associations with human health and disease. Studies demonstrate that gut microbes play critical roles in digestion, nutrient absorption, and immune system modulation \citep{jandhyala2015role, valdes2018role}. 
These findings underscore the potential of gut microbiome research for therapeutic and preventive applications.
However, despite these promising associations, their translational impact remains limited, as the underlying mechanisms driving host-microbiome interactions are still poorly understood. Without a mechanistic framework, it remains challenging to develop targeted interventions or accurately predict disease outcomes based on microbiome composition.

Since the integrative Human Microbiome Project (iHMP), multiomics studies—including genomics, transcriptomics, proteomics, and metabolomics—have advanced our understanding of the mechanism underlying the host-microbiome interactions \citep{jiang2019microbiome, integrative2019integrative_new}. Among these approaches, paired microbiome-metabolome studies (PM2S) provide unique insights into gut metabolism, a key mechanism by which the gut microbiome influences host health \citep{muller2022gut}. 
By simultaneously profiling hundreds of gut microbes and metabolites, PM2S enables the systematic identification of microbial drivers of metabolism, capturing complex interactions that would otherwise remain undetected. 
Understanding microbiome-metabolome interactions is critical for bridging the gap between microbiome research and clinical applications, as metabolic byproducts often serve as direct mediators of host physiology. 

A key challenge of analyzing microbiome-metabolome interactions lies in the complex dependencies among microbes. As a result, identifying true microbial drivers for each gut metabolite requires multivariate analysis of hundreds of correlated microbial features—an inherently high-dimensional task. 
This complexity, combined with limited sample sizes, severely constrains statistical power.
PM2S studies are still relatively scarce due to the novelty and high cost of simultaneously profiling both microbiomes and metabolomes. As such, existing datasets are often small in scale; for instance, the IBD cohort in the iHMP includes only 132 subjects, despite profiling thousands of microbial and metabolic features \citep{integrative2019integrative_new}. Biological weak signals further exacerbate this issue, as gut metabolite production frequently involves subtle, distributed contributions from multiple microbial taxa and their interactions with host physiology \citep{liu2022functions}.
These challenges underscore the need for advanced statistical methods that can accommodate high dimensionality, account for microbial dependencies, and extract weak signals from limited PM2S data. 

A potential strategy to address the low-power issue in PM2S studies is to leverage auxiliary data through transfer learning. When the auxiliary data are informative and well-aligned with the primary task, transfer learning can significantly enhance predictive performance \citep{zhuang2020comprehensive}.
However, existing transfer learning methods are not readily applicable to PM2S analyses for several important reasons. 
First, most transfer learning approaches are designed to improve prediction accuracy, whereas the primary goal in PM2S studies is inference—specifically, identifying microbiome-metabolome interactions. 
Second, PM2S datasets are not only scarce but also heterogeneous in both microbiome and metabolome measurements. 
More specifically, different studies may use distinct sequencing technologies (e.g., targeting different hypervariable regions of the 16S rRNA gene) and preprocessing pipelines (e.g., Fastp \citep{chen2018estimating} vs. Bowtie2 \citep{langmead2012fast}), leading to inconsistent microbial feature spaces. 
On the metabolomics side, differences in platforms (e.g., NMR vs. mass spectrometry) and quantification methods result in varying coverage, sensitivity, and specificity for metabolite detection.
These inconsistencies make it difficult to find truly transferable external PM2S datasets and raise the risk of introducing bias through negative transfer. 

\subsection{Related Work}


Current analyses of microbiome-metabolome interactions using PM2S data often rely on univariate tests or simple regression models \citep{hooper2012interactions, visconti2019interplay}, which fail to capture the complex dependencies among microbes and metabolites, increasing the risk of false discoveries.
To address these limitations, multivariate methods have been increasingly adopted. For example, canonical correlation analysis identifies groups of microbes and metabolites with strong interactions \citep{nguyen2021associations}, while high-dimensional regression models like lasso and elastic net detect associations between gut metabolites and microbial communities \citep{mallick2019predictive}.  Recently, MiMeNet \citep{reiman2021mimenet} introduced a neural network framework to model complex, nonlinear relationships between microbiome and metabolome profiles. 
Additionally, approaches tailored to microbiome data have emerged, such as regression with compositional covariates to account for microbial relative abundance \citep{xia2013logistic, shi2016regression} and kernel-based regression models that incorporate phylogenetic relationships to improve statistical power and biological relevance \citep{zhao2015testing, randolph2018kernel, wang2023generalized}.

Our work is closely related to a growing line of research on transfer learning, which aims to leverage auxiliary (source) data to improve prediction or estimation in a target domain. For comprehensive overviews, we refer readers to several survey articles \citep{pan2009survey, weiss2016survey, zhuang2020comprehensive, NLW20} and the references therein. Beyond algorithmic developments, recent efforts have focused on establishing the statistical properties of transfer learning methods within specific modeling frameworks.
For example, \citet{Bastani2021} studied transfer learning in high-dimensional linear models with a single informative auxiliary dataset, assuming the auxiliary sample size exceeds the number of covariates. \citet{LCL21} introduced the trans-Lasso, a method that accommodates multiple informative auxiliary datasets to improve estimation in high-dimensional linear models, and this framework was later extended to generalized linear models (GLMs) by \citet{tian2023transfer}. 
\cite{gu2024robust} relaxes the assumption in these high-dimensional regression works by adopting an angle-based similarity measure instead of a distance-based one.
\citet{cai2021transfer} proposed a transfer learning approach for high-dimensional Gaussian graphical models, demonstrating faster convergence rates than those attainable with a single study.
For broader function classes, \citet{tripuraneni2021provable} developed theoretically grounded meta-learning techniques to estimate multiple related models sharing a low-dimensional representation.
Recent contributions such as TransHDGLM by \citet{Li02042024} provide rate-optimal estimators with provable asymptotic guarantees for high-dimensional GLMs, while \citet{Guo18022025} developed a federated meta-learning inference framework with applications to multi-site EHR data from COVID-19 patients.


Existing transfer learning methods typically rely on the assumption that each auxiliary (source) dataset is individually informative of the target. 
However, this requirement poses significant challenges in microbiome-metabolome interaction analysis for two key reasons. First, informative external datasets are scarce due to the cost of PM2S experiments and the inherent heterogeneity of microbial and metabolic profiles. 
Second, weak biological signals, high noise levels, and the concerted nature of microbe-metabolite interactions make it unlikely that any single auxiliary dataset will be strongly informative of the target. Instead, the collective combination of these datasets may provide joint informativeness—a scenario underexplored in current transfer learning frameworks.



\subsection{Contributions}\label{sec:contribution}
The primary contribution of our project is {the development of CoMMiT, a within-cohort transfer-learning framework} designed to detect weak microbiome-metabolome interactions in studies with limited sample sizes. 
Unlike existing transfer-learning approaches, our framework transfers knowledge within a single PM2S dataset without using external datasets. 
This design also minimizes the risk of negative transfer that may arise from the substantial data heterogeneity across PM2S studies. 
CoMMiT is built on a novel projection-based learning mechanism, which leverages shared information from auxiliary metabolites, even when these metabolites are only weakly informative. 
This flexibility is essential for real-world microbiome applications, where strong signals are rare due to noise.
Notably, {CoMMiT will also enable a more accurate prediction of target metabolite abundance for metagenomic samples without corresponding metabolomic data.} 
This capability is particularly valuable in microbiome research, where the budget limitations often result in a lack of paired metabolomic data in human microbiome studies. 
Even for studies that do include paired microbiome-metabolome data, this prediction tool remains valuable. For instance, in targeted metabolomics studies, the metabolite panels may be limited to specific targets, leaving other metabolites unmeasured. In untargeted studies, issues such as resolution limitations can result in substantial missing data for metabolite measurements.

CoMMiT is theoretically sound: The CoMMiT estimator is consistent with an established convergence rate, and a de-biased CoMMiT estimator is further developed to enable efficient $p$-value calculation for identifying microbiome-metabolome interactions. 
We also provide a data-driven procedure for selecting auxiliary metabolites to achieve optimal performance of estimation and inference. 
Extensive simulation studies and a comprehensive application to the CARB study demonstrate CoMMiT's effectiveness in predicting metabolites and identifying biologically meaningful gut microbes associated with bile acids.

\subsection{Notation}

Throughout the paper, we use normal typeface to denote scalars, bold lowercase typeface to denote vectors and bold uppercase typeface to denote matrices.  For any vector ${\bf v} \in \mathbb{R}^p$, we use $v_l$ to denote the $l$-{th} element of $\bf v$ for $l = 1, \ldots, p$. For any matrix ${\bf M} \in \mathbb{R}^{n \times p}$, let $ {\bf m}_l$ and $m_{il}$, respectively, denote the $l$-{th} column and $(i,l)$ entry of $\bf M$ for $i = 1, \ldots, n$ and $l = 1, \ldots, p$. For any index set $\mathcal{I} \subset \{1, \ldots, p\}$, let ${\bf v}_\mathcal{I}$ and ${\bf M}_{\mathcal{I}}$ denote the subvector of ${\bf v}$ whose elements are indexed by $\mathcal{I}$ and the submatrix of ${\bf M}$ whose columns are indexed by $\mathcal{I}$, respectively. Let $I(\mathcal{A})$ denote the indicator function of the event $\mathcal{A}$; i.e., $I(\mathcal{A}) = 1$ if $\mathcal{A}$ is true, and $I(\mathcal{A}) = 0$ otherwise. Finally, we let $ \|{\bf v}\|_0 := \sum_{{l} = 1}^p I(v_l \neq 0), ~~ \|{\bf v}\|^q_q := \sum_{j = 1}^p |v|^q $ for any $ 0 < q < \infty$, $\|{\bf v}\|_{\infty} := \max_l|v_l|$,  $\|{\bf M}\|_q := \sup_{\|{\bf v}\|_q = 1}\|{\bf Mv}\|_q$ for any $q > 0$ and $\|{\bf M}\|_F^2 := \sum_{i=1}^n \sum_{l=1}^p m_{ij}^2$.  We also denote $\text{span}({\bf v}_1, \ldots, {\bf v}_p)$ as the linear space spanned by ${\bf v}_1, \ldots, {\bf v}_p$.

\section{Data Description} \label{sec:carb}
The specific study that motivated this paper is the ``Carbohydrates and Related Biomarkers" (CARB) study, a randomized, controlled, crossover feeding study at the Fred Hutchinson Cancer Center aimed at evaluating the effects of low and high glycemic load on various biomarkers, such as systemic inflammation, insulin resistance, and adipokines \citep{neuhouser2012low}. 
Participants, randomized based on body mass index and sex, were fed two controlled eucaloric diets with differing glycemic loads for 28 days, separated by a 28-day washout period. 
Participants collected stool samples at the end of each 28 day period.  
The stool samples were used for the 16S rRNA gene data and 
the targeted metabolomics profiling of bile acids \citep{ginos2018circulating}.

Bacterial taxonomy was classified using QIIME \citep{caporaso2010qiime}. After quality control (QC), we obtained $8.3 \times 10^7$ sequences, averaging $2.3 \times 10^5$ sequences per sample with an average length of 290 bp. This analysis identified 9 bacterial phyla, 1 Archaea, and 161 genera. Filtering further reduced the number to 134 bacterial genera present in at least 15\% of participants.
The microbial abundances were further transformed using the centered-log-ratio (CLR) transformation \citep{aitchison1982statistical}.  
Quantitative targeted metabolomics profiling of 55 bile acids was performed using liquid chromatography tandem mass spectrometry (LC-MS/MS).
Specifically, standard operating procedures from the Northwest Metabolomics Research Center were followed,  
as detailed in a previous study \citep{ginos2018circulating}. 
Using multiple-reaction-monitoring (MRM) mode, targeted data
were acquired for 55 bile acids. 
All 55 bile acids and five stable isotope-labeled internal
standards MRM transitions were monitored in the negative
mode. Metabolite identities were validated by spiking mixtures
of standard compounds. 
TargetLynx software (Waters, Milford,
MA) was used to integrate extracted MRM peaks. Absolute bile
acid concentrations were calculated
from the peak areas in the samples and the slope of the calibration curve \citep{raftery2016mass, sarafian2015bile}. 
Twenty-six blinded quality control samples were used to determine the coefficient of variation (CV) for each bile acid, which ranged from 9\% to 150\%. Bile acids with very low signal or CVs $> 30\%$ were excluded. 
As a result, of the 55 bile acids measured, 29 were retained for analysis. 
We log-transformed the abundance of these bile acids to account for skewness.

\begin{figure}[!ht]
	\includegraphics[width=0.5\textwidth, height=5cm]{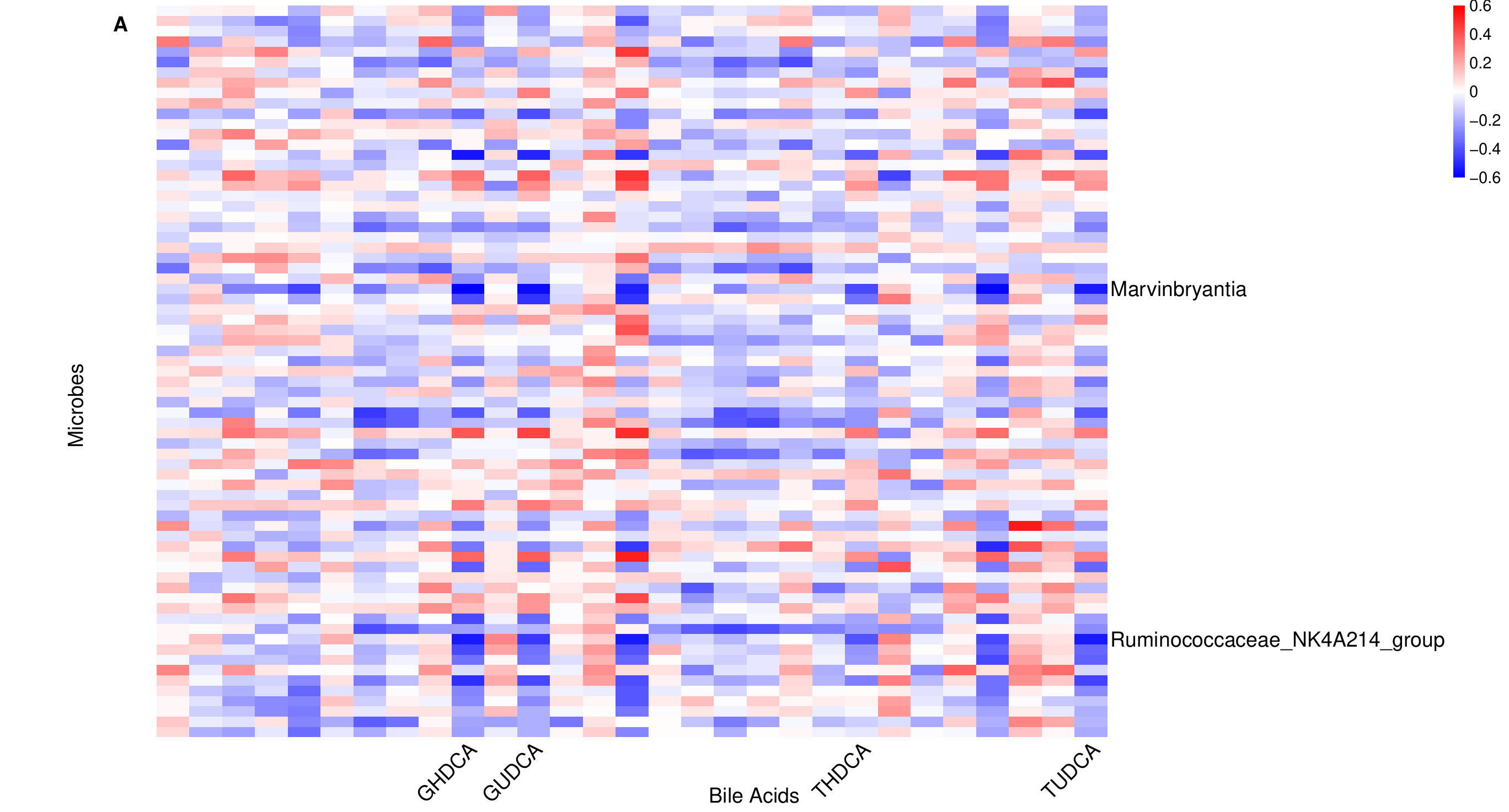}
    \includegraphics[width=0.5\textwidth, height=5cm]{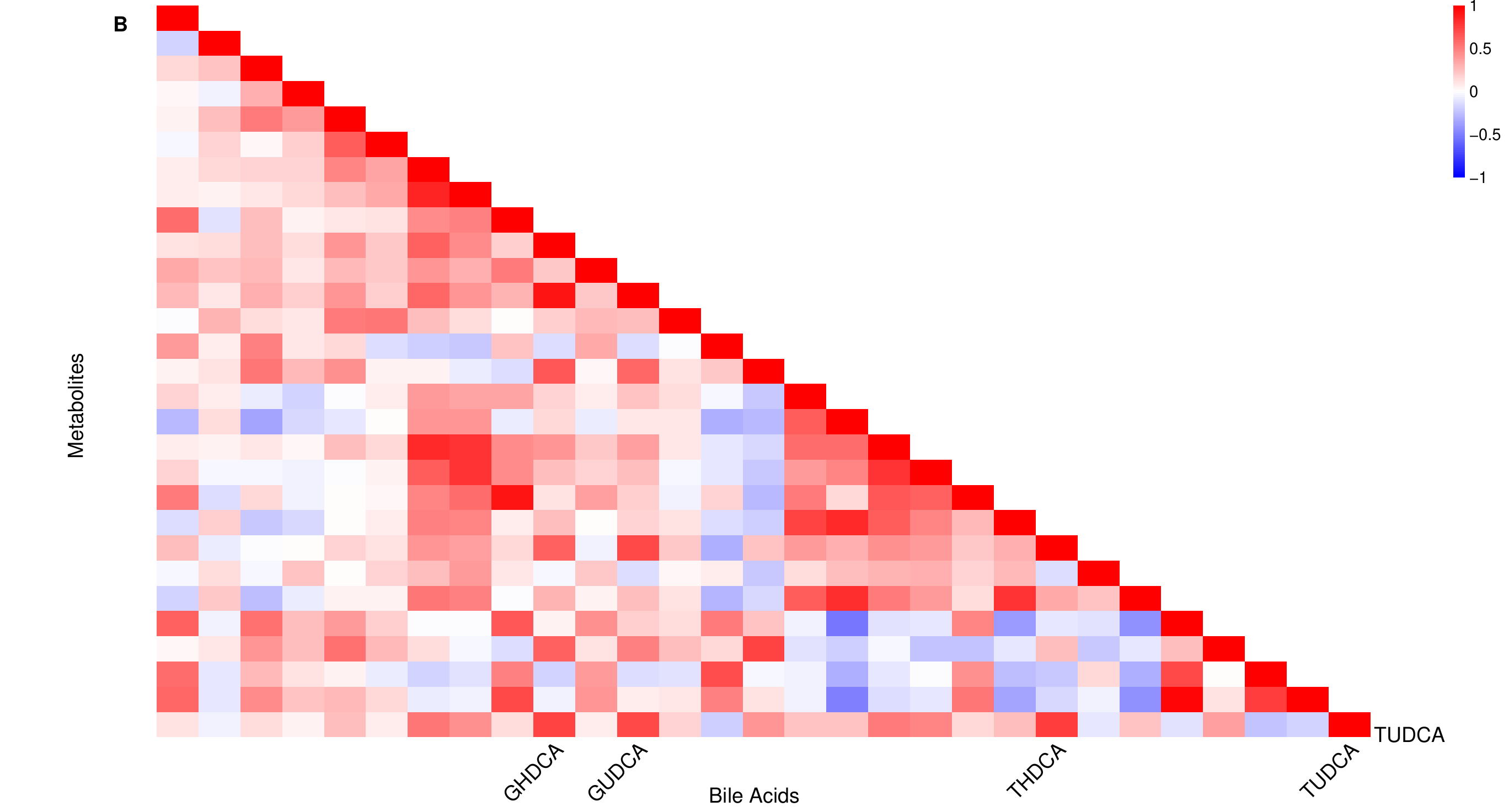}
	\caption{(A) Microbiome-metabolome correlation heatmap: displays the correlations between $134$ gut microbes and $29$ bile acids. For better visual display, we only show 71 gut microbes. Each row represents a distinct gut microbe, and each column corresponds to a different bile acid, with TUDCA displayed in the last column. (B) Metabolome-wide correlation heatmap: each row and column represents one of the $29$ bile acids, 
 including TUDCA. 
 The off-diagonal entries show correlation estimates, with a value of 1 on the diagonal indicating the correlation between each bile acid and itself. Both heatmaps exhibit interactions between TUDCA and gut microbes, as well as correlations between TUDCA and other metabolites.}
	\label{Fig1}
\end{figure}

We hypothesized that the interplay between gut microbes and bile acids, i.e., metabolites critical for fat and fat-soluble vitamin metabolism, represents a key mechanistic pathway through which the low-glycemic-load (LGL) diet exerts its microbiota-mediated health benefits.
As a preliminary analysis to examine this hypothesis, Fig. \ref{Fig1}A displays the overall patterns of microbiome-bile acid correlations, highlighting the interaction between various bile acids and gut microbes. 
One notable metabolite is {tauro-ursodeoxycholic acid (TUDCA)}, shown in the last column of Fig. \ref{Fig1}A. 
TUDCA, formed from a secondary bile acid {ursodeoxycholic acid (UDCA)} in the liver, is known for potential health benefits, including liver protection and reducing the risk of diseases like Huntington's disease \citep{khalaf2022tauroursodeoxycholic}, as well as obesity and stroke \citep{vang2014unexpected}. 
TUDCA shows marginal correlations with various gut microbes, such as {Marvinbryantia} and {Ruminococcaceae NK4A214 group}, but these correlations may be confounded by complex microbial dependencies \citep{song2019taxonomic}. 
While a high-dimensional model can theoretically account for these dependencies,  
with a limited sample size ($n=73$) and a large number of microbes ($p=134$), we anticipate a low statistical power and potentially high false discovery rate. 
CoMMiT will enhance the statistical power by leveraging information from other bile acids that are correlated with TUDCA through shared microbial pathways. 
As demonstrated in Fig.\ref{Fig1}B, secondary bile acid conjugates like {glycohyodeoxycholic acid (GHDCA)}, {glycoursodeoxycholic acid (GUDCA)}, and {taurohyodeoxycholic acid (THDCA)} show strong correlations with TUDCA. 
Furthermore, Fig. \ref{Fig1}A shows that they also share similar correlation patterns with the microbes.

These preliminary data suggest the scientific feasibility of leveraging co-informative metabolites within a single PM2S study for more powerful detection of microbiome-metabolome interactions, which drive the methodological development outlined in Section \ref{sec:algorithm}. 

\section{Method}\label{sec:algorithm}
\subsection{The CoMMiT model}\label{subsec:model}

We denote ${\bf x} \in \mathbb{R}^p$ as the gut microbiome data after taking the centered log-ratio transformation \citep{aitchison1982statistical}. 
Letting $y^{(j)} \in \mathbb{R}$ denote the $j$-th metabolite abundance (after log-transformation) for $j = 0, \ldots, J$, 
with $y^{(0)}$ denoting the target metabolite and $y^{(1)}, \ldots, y^{(J)}$ denoting auxiliary metabolites that are informative for $y^{(0)}$. 
We consider the target model
\begin{align}\label{eq:0}
	{y}^{(0)} = {\bf x}^\intercal \ve \beta^{(0)} + \varepsilon^{(0)},  
\end{align}
and the auxiliary models 
\begin{align}\label{eq:0:aux}
	{y}^{(j)} = {\bf x}^\intercal \ve \beta^{(j)} + \varepsilon^{(j)} \mbox{ for } j=1,\ldots, J, 
\end{align}
where $\mathbb{E}[\varepsilon^{(j)} \mid {\bf x} ] = {0}$ and $\mathbb{E}[(\varepsilon^{(j)})^2 \mid {\bf x}] = \sigma_j^2$ for $j = 0, 1, \ldots, J$.
Within this framework, 
CoMMiT assumes the following projection-based similarity:
\begin{align}\label{wang:ye:1}
	\left\|\ve \beta^{(0)} - \sum_{j=1}^J \alpha_j \ve \beta^{(j)} \right\|_1 \leq h, 
\end{align}
for some $h>0$ and coefficients $\alpha_1, \ldots, \alpha_J$. 

\begin{figure}[ht!]
    \centering
    \includegraphics[scale=0.47]{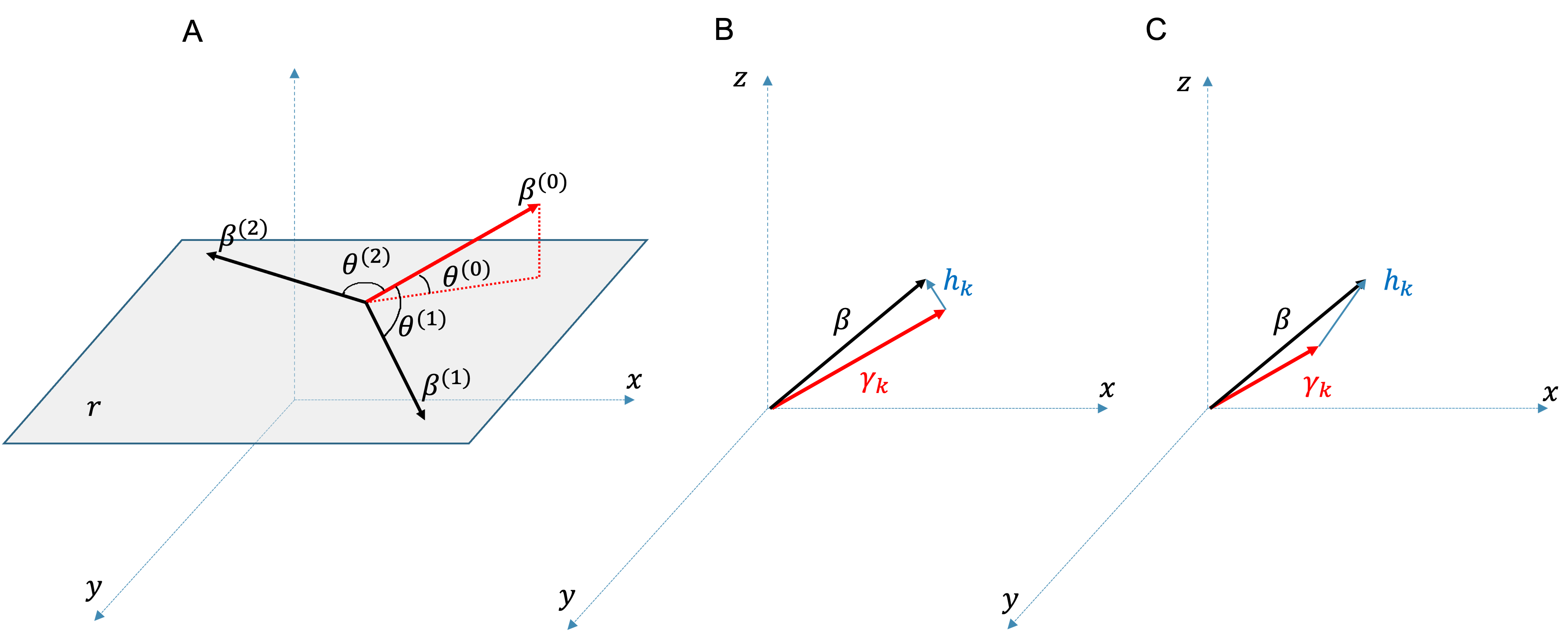}
    \caption{(A): An illustration of the proposed projection-based similarity with $J=2$: the coefficient of interest $\ve \beta^{(0)}$ is aligned with the space spanned by $\ve \beta^{(1)}$ and $\ve \beta^{(2)}$, but not aligned with either $\ve \beta^{(1)}$ or $\ve \beta^{(2)}$. (B): Distance-based similarity \citep{li2022transfer}. (C): Angle-based similarity \citep{gu2024robust}.}
    \label{Fig2}
\end{figure}

Fig. \ref{Fig2}A illustrates our projection-based similarity in the Euclidean space with an example involving $J = 2$ auxiliary metabolites, where $\theta^{(0)}$ refers to the angle between $\ve \beta^{(0)}$ and the two-dimensional space spanned by $\ve \beta^{(1)}$ and $\ve \beta^{(2)}$, and $\theta^{(j)}$ refers to the angle between $\ve \beta^{(0)}$ and $\ve \beta^{(j)}$ for $j = 1,2$. 
Notably, while $\theta^{(0)}$ is small, both $\theta^{(1)}$ and $\theta^{(2)}$ are large. 
This example underscores the flexibility of CoMMiT: CoMMiT does not require any of the auxiliary metabolites to be individually informative of the target metabolite, which, however, is the assumption adopted in existing transfer-learning methods. 
For example, as shown in Fig. \ref{Fig2}B, \cite{li2022transfer} and \cite{tian2022transfer} 
both assume the informativeness of each auxiliary data source through a distance-based measure, i.e., 
\begin{align}\label{li:1}
	\left\|\ve \beta^{(0)} - \ve \beta^{(j)}\right\|_q \leq h \mbox{ for some } q \in [0,1] 
\end{align}
for all $j = 1, \ldots, J$. 
\cite{gu2024robust} requires the angle between $\ve \beta^{(0)}$ and each $\ve \beta^{(j)}$ to be small (Fig. \ref{Fig2}C); that is, 
\begin{align}\label{gu:1}
	\left| 1- \frac{ \left({\ve \beta^{(0)}}^\intercal \ve \beta^{(j)}\right)^2 }{ \|\ve \beta^{(0)}\|^2 \|\ve \beta^{(j)}\|^2 } \right| \leq h.
\end{align}
While \eqref{gu:1} relaxes \eqref{li:1} by removing the scale differences, it still requires each $\ve \beta^{(j)}$ to be informative for the target $\ve \beta^{(0)}$.
In contrast, our projection-based similarity (\ref{wang:ye:1}) allows substantially more flexibility.  
On one hand, when none of the individual auxiliary metabolites satisfy \eqref{li:1} or \eqref{gu:1}, 
they can still satisfy \eqref{wang:ye:1} by collectively informing the target metabolite.
On the other hand, if \eqref{li:1} or \eqref{gu:1} is satisfied,  
then it is easy to verify that our assumption (\ref{wang:ye:1}) also holds.

\subsection{The CoMMiT Estimation}\label{subsec:estimation}
Recall from \eqref{eq:0} and \eqref{eq:0:aux} that ${y}^{(j)} = {\bf x}^\intercal \ve \beta^{(j)} + \varepsilon^{(j)}$ with $\mathbb{E}[\varepsilon^{(j)} \mid {\bf x}^\intercal] = 0$ and $\mathbb{E}[ (\varepsilon^{(j)})^2 \mid {\bf x}^\intercal] = \sigma_j^2$
for $j=0, 1, \ldots, J$, 
where $y^{(0)}$ is the target metabolite, and $y^{(1)}, \ldots, y^{(J)}$ are the auxiliary metabolites from the same cohort. 
The CoMMiT estimation is a high-dimensional estimation procedure for the target parameter $\ve \beta^{(0)}$ that efficiently leverages information from the auxiliary parameters $\ve \beta^{(1)}, \ldots, \ve \beta^{(J)}$. Motivated by the proposed projection-based similarity (\ref{wang:ye:1}), we write
\begin{equation}\label{beta:decom:2}
	\ve \beta^{(0)} = \sum_{j=1}^J \alpha_j \ve \beta^{(j)} + {\ve w},
\end{equation}
indicating that one can estimate $\ve \beta^{(0)}$ by separately estimating $\{\alpha_j\}_{j=1}^J, \{\ve \beta^{(j)}\}_{j=1}^J$, and $\ve w$. 

Let $\{y_i^{(0)}, \ldots, y_i^{(J)}, {\bf x}_i\}_{i=1}^n$ denote $n$ $i.i.d$ realizations of $\{y^{(0)}, \ldots, y^{(J)}, {\bf x}\}$. The CoMMiT estimation procedure consists of four steps. 

\begin{enumerate}
	\item For each $j=1,\ldots,J$, find  \begin{equation}\label{step1}
		\widehat{\ve \beta}^{(j)} = \mbox{argmin}_\ve \beta \left\{ \frac{1}{2n} \sum_{i=1}^n \left(y_i^{(j)} - {\bf x}_i^\intercal \ve \beta \right)^2 + \lambda_j\|\ve \beta\|_1\right\},
	\end{equation}
	where $\lambda_j = c_j\sqrt{\log p/n}$ for some positive constants $c_j>0$, $j=1,...,J$. 
	\item Compute 
	\begin{equation}\label{step2}
		\widehat{\ve \alpha},\widehat{\ve w}=\mbox{argmin}_{\ve \alpha,\ve w} \left\{ \frac{1}{2n} \sum_{i = 1}^n \left(y_i^{(0)} - \sum_{j=1}^J \alpha_j {\bf x}_i^\intercal \widehat{\ve \beta}^{(j)}  -
 {\bf x}_i^\intercal\ve w \right)^2 + \lambda_w\|\ve w\|_1
        \right \},
	\end{equation}
	where $\lambda_w=c \sqrt{{\log p}/{n}} $ for some positive constant $c$.
	\item Calculate the CoMMiT estimator
	\begin{equation}\label{step4}
		\widehat{\ve \beta}^{(0)} = \sum_{j=1}^J \widehat{\alpha}_j \widehat{\ve \beta}^{(j)} + \widehat{\ve w}.
	\end{equation}
\end{enumerate}
Step 1 finds the Lasso estimate for each $\ve \beta^{(j)}$. Step 2 estimates $\ve \alpha$ and $\ve w$ simultaneously with an $l_1$-penalty on $\ve w$.
We require the following assumptions for theoretical analyses $\widehat{\ve \beta}^{(0)}$. 
\begin{description}
\item [{A}] (Sparsity) 
There exist constants $s_w$ and $s_j$ such that $\|\boldsymbol{w}\|_{0}\le s_{w}$,
 $\|\bbeta^{(j)}\|_{0}\le s_{j}$ for $j=1,...,J$, and $\sstar:=\max_{j=1,...,J}s_{j}<\infty$.

\item[{B}] (Sparse Reisz Condition (SRC))
There exist constants $\underline{c},\bar{c}>0$
(not depending on $n$) such that for any $\bbeta \in \mathbb{R}^p$
with $\left\Vert \bbeta\right\Vert _{0}\le\min(\max\{s^*,s_w\},\text{p})$,
we have 


\begin{equation}
\underline{c}\left\Vert \bbeta\right\Vert _{2}\leq\frac{1}{\sqrt{n}}\left\Vert \mathbf{X}\bbeta\right\Vert _{2}\leq \bar{c}\left\Vert \bbeta\right\Vert _{2},\label{eq:src}
\end{equation}
where ${\bf X} = ({\bf x}_1, \ldots, {\bf x}_n)^\intercal$. 
\end{description}

\begin{remark}
    The SRC assumption was first proposed in \cite{zhang2008sparsity}. It requires that the eigenvalues of  $\frac{1}{n}\mathbf{Z}^T\mathbf{Z}$ are bounded above and away from 0, for any low-dimensional sub-matrix $\mathbf{Z}$ of $\mathbf{X}$. Such an assumption is standard in the $l_{1}$-regularization analysis. Similar assumptions in the literature include sparse eigenvalue conditions in \cite{zhang2010analysis,raskutti2011minimax}, the quasi-isometry condition in \cite{Rigollet:2011fz} and the more stringent ($\underline{c}$, $\bar{c}$ close to 1) restricted isometry property in \cite{candes2007dantzig}.

\end{remark}

The following theorem establishes the convergence rate of $\widehat{\ve \beta}^{(0)}$. 
\begin{theorem}\label{thm1}
Consider the linear models in \eqref{eq:0} and \eqref{eq:0:aux}. 
Under Assumptions A and  B, we have the following $$\left\|\widehat{\ve \beta}^{(0)}-\ve \beta^{(0)}\right \|_{2}^{2}=O_{p}(J\sstar\frac{\log p}{n} + \sqrt{\frac{\log p}{n}} h\land h^{2}),$$ 
where $h$ is defined in \eqref{wang:ye:1} , and $J$ is the total number of auxiliary outcomes.  
\end{theorem}

{Theorem \ref{thm1} highlights an interesting trade-off phenomenon between 
$J$ and $h$. When $J$ increases, that is, more auxiliary metabolites are incorporated, 
we know $h$ will likely decrease. 
In this scenario, the accuracy of the CoMMiT estimator 
$\widehat{\bm{\beta}}^{(0)}$ depends on the rate at which 
$h$ decreases relative to the increase in $J$. 
This observation also underscores the importance of selecting the most informative set of auxiliary metabolites, a topic we will explore further in Section \ref{sec:heuristic}.
}

\begin{remark}
    As we can see from Theorem \ref{thm1}, when $J$ is fixed, and $h = o( s^*\sqrt{\log p/n} )$, 
    our CoMMiT estimator has the convergence rate of order $s^* \cdot \frac{\log p}{n}$. This rate is comparable to the classic rate of LASSO estimator $s\cdot\frac{\log p}{n}$. 
      Note that a faster convergence rate was shown in existing transfer-learning literature such as \cite{li2020transfer}. 
    This is because those studies leverage auxiliary samples, thereby increasing the effective sample size. 
    However, since we transfer information within the same cohort, the faster convergence rate is not attainable.     
\end{remark}

\subsection{The CoMMiT Inference}\label{sec:inference}
In this section, we introduce the CoMMiT inference procedure
to construct an asymptotic test statistic for testing $H_{0,l}: \beta^{(0)}_l = 0$ for some $l = 1, \ldots, p$, which can be used to identify which microbes are associated with the target metabolite. 
Note that 
the proposed CoMMiT estimator cannot be directly used for inference due to the well-known bias of $l$-1 penalized estimators. 
To overcome the difficulty, we correct the bias of $\widehat{\ve \beta}^{(0)}$ following the general idea of debiasing for lasso-type estimators
\citep{vandegeer2014, zhangzhang2014}. 
Letting ${\bf x}_{i, -l}$ denote the subvector of ${\bf x}_i$ with the $l$-th element excluded, we regress $x_{il}$ on ${\bf x}_{i, -l}$ and obtain the residual 
\begin{equation}\label{residual}
{z}_{il} = {x}_{il} - {\bf x}_{i, -l}^\intercal\widehat{\ve \gamma}^{(l)}, 
\end{equation}
where
\[
\widehat{\ve \gamma}^{(l)} = \argminB_{\ve \gamma}  \left\{  \frac{1}{2n} \sum_{i=1}^n \left({x}_{il} - {\bf x}_{i, -l}^\intercal\ve \gamma\right)_2^2 + \zeta_l \|\ve \gamma\|_1 \right\}
\]
with the tuning parameter $\zeta_l > 0$ for $l = 1, \ldots, p$. Then,
we calculate the bias-corrected estimator for the $j$-th auxiliary regression coefficient
$\ve \beta^{(j)}$, denoted by $\widehat{\ve \beta}^{(j)}_{\text{de}} = \left( \widehat{ \beta}^{(j)}_{1, \text{de}}, \ldots, \widehat{ \beta}^{(j)}_{p, \text{de}} \right)^\intercal$, where
\begin{equation}
  \widehat{\beta}^{(j)}_{l, \text{de}} = \widehat{\beta}^{(j)}_l + \frac{\sum_{i=1}^n { z}_{il} \left({y}_i^{(j)} - {\bf x}_i^\intercal \widehat{\ve \beta}^{(j)}\right)}{\sum_{i=1}^n { z}_{il} {x}_{il}}\label{eq:app_error}  
\end{equation}
and $\widehat{\bm{\beta}}^{(j)}$ is defined in \eqref{step1} for $l = 1, \ldots, p$ and $j = 1, \ldots, J$. 
Similarly, we construct the bias-corrected estimator of $\ve w$ as
\[
\widehat{\ve w}_{\text{de}} = \left( \widehat{ w}_{1,\text{de}}, \ldots, \widehat{ w}_{p,\text{de}} \right)^\intercal, ~~~ \widehat{ w}_{l,\text{de}} = \widehat{w}_l + \frac{\sum_{i=1}^n { z}_{il} \left({r}_i^{(0)} - {\bf x}_i^\intercal \widehat{\ve w}\right)}{\sum_{i=1}^n {z}_{il} {x}_{il}} \mbox{ for } l = 1, \ldots, p,
\]
where $r_i^{(0)} = y_i^{(0)} - \sum_{j=1}^J \widehat{\alpha}_j \widehat{\ve \beta}^{(j)}$ with  
 $\widehat{\alpha}_j$ and $\widehat{\ve w}$ defined in \eqref{step2}. 
Finally, following (\ref{beta:decom:2}), 
we define our de-biased estimator of $\ve \beta^{(0)}$ as 
\begin{align}\label{Lasso-AO:de}
\widehat{\ve \beta}^{(0)}_{\text{de}} = \sum_{j=1}^J \widehat{\alpha}_j \widehat{\ve \beta}^{(j)}_{\text{de}}+ \widehat{\ve w}_{\text{de}}.
\end{align}

We next establish the asymptotic distribution of $\widehat{\ve \beta}^{(0)}_{\text{de}}$. We need some additional conditions. 

\begin{description}
\item[C]{(Bounded Bias Factor)} Let $X=({\bf x}_1,...,{\bf x}_n)^T$ denote the design matrix. Denote the $k$-th column of $X$ as $X_{\cdot k}$. Assume that $\eta^{*}:=\max_{l=1,...,p}\eta_{l}<C_\eta$ for some constant $C_\eta$, where $\eta_{l}=\max_{k\neq l}|{\bf z}_{l}^{\intercal}X_{\cdot k}|/\|{\bf z}_{l}\|_{2}$  with ${\bf z}_l = (z_{1l}, \ldots, z_{nl})^\intercal$ for $l=1,...,p$ defined in \eqref{residual}.

\end{description}
\begin{remark}
{ 
In \cite{zhangzhang2014}, $\eta_l$ is referred to as the {\it bias factor} due to the fact that $\max_{k\neq l}|{\bf z}_{l}^{\intercal}{X}_{\cdot k}|$ controls the term $\sum_{i=1}^n { z}_{il} \left({y}_i^{(j)} - {\bf x}_i^\intercal \widehat{\ve \beta}^{(j)}\right)$ in \eqref{eq:app_error}
for $j=0,1,...,J$.}
\end{remark}

\begin{theorem}
Define $\tau_{l}={|{\bf z}_{l}^{\intercal}{\bf x}_{l}}|/{\|{\bf z}_{l}\|_{2}}$. 
Under Conditions A, B and C,
if $h\le c_{1}s^{*}\sqrt{\frac{2}{n}\log\frac{p}{\epsilon}}$ for some positive constant $c_1$, we have 
\[
P(|\widehat{\beta}_{l,\textrm{de}}^{(0)}-\beta_{l}^{(0)}|\ge t)\le2F_{t_{n}}(- t\tau_{l}/\widehat{\sigma}_{0}),
\]
as $n \rightarrow \infty$, where $\hat{\sigma}_0$ is a consistent estimator of $\sigma_0$ that will be discussed later. In particular, this leads to a $100(1-\alpha)\%$ confidence interval
for $\beta_{l}^{(0)}$: 
\[
\widehat{\beta}_{l,\textrm{de}}^{(0)}\pm t_{n,1-\alpha/2}\cdot \widehat{\sigma}_{0}/\tau_{l}.
\]
\end{theorem}
Finally, we discuss how to estimate $\sigma_0^2$. 
A naive estimator of $\sigma_0^2$ is the sum-of-squares type estimator:
\[
\widehat{\sigma}_{0, \text{naive}}^2 = n^{-1}\sum_{i=1}^n \left({y}_i^{(0)} - {\bf x}_i^\intercal \widehat{\ve \beta}^{(0)}_{\text{de}}\right)^2. 
\]
However, in the classical $n > p$ setting where $\widehat{\ve \beta}^{(0)}_{\text{de}}$ reduces to the least-squares estimator, 
some calculation shows that such an estimator is biased downward, 
leading to potentially inflated type-I error rates. We refer readers to \cite{yu2019estimating} for more details.
Recent literature corrects this bias by constructing various modified estimators. For example, \cite{sunzhang2012} and \cite{fan2012variance} both studied the estimator
\[
\widehat{\sigma}_{0,\text{sf}}^2 = (n - \widehat{s})^{-1} \sum_{i=1}^n \left({y}_i^{(0)} - {\bf x}_i^\intercal \widehat{\ve \beta}^{(0)}\right)^2,
\]
where $\widehat{s}$ is the number of nonzero elements in $\widehat{\ve \beta}^{(0)}$. The performance of $\widehat{\sigma}_{0,\text{sf}}^2$ depends on the magnitude of the set of nonzero elements in $\ve \beta^{(0)}$. However, when $\ve \beta^{(0)}$ has many non-zero entries, which is theoretically possible under our assumptions, 
we may have $\widehat{s} \approx n$, leading to a large value of $\widehat{\sigma}_{0,\text{sf}}^2$ and low statistical power. 
Thereby, we adopt the natural lasso estimator \citep{guo2019optimal}, which addresses the issues of 
$\widehat{\sigma}^2_{0, \text{naive}}$ and $\widehat{\sigma}^2_{0, \text{sf}}$:
\[
\widehat{\sigma}_{0, \text{nat}}^2 = n^{-1} \sum_{i=1}^n \left({y}_i^{(0)} - {\bf x}_i^\intercal \widehat{\ve \beta}_{\text{nat}} \right)^2 + 2\lambda_{\text{nat}} \|\widehat{\ve \beta}_{\text{nat}}\|_1,
\]
where\[
\widehat{\ve \beta}_{\text{nat}} = \argminB_{\ve \beta} \left\{ n^{-1} \left({y}_i^{(0)} - {\bf x}_i^\intercal {\ve \beta} \right)^2 + 2\lambda_{\text{nat}} \|\ve \beta\|_1 \right\}.
\]
As pointed out in \cite{yu2019estimating}, the consistency of $\widehat{\sigma}_{0, \text{nat}}^2$ requires that $\|\ve \beta^{(0)}\|_1 = o(\sqrt{n/\log p})$, which is satisfied if $s^* = o(\sqrt{n/\log p})$ and $s_w = o(\sqrt{n/\log p})$.


\subsection{Heuristic selection of the informative metabolite set}\label{sec:heuristic}
In the CoMMiT estimation and inference procedures, we assume the auxiliary metabolites satisfying \eqref{wang:ye:1} are pre-specified. 
In practice, an initial pool of such metabolites can be obtained from metabolic pathways in existing metabolic databases such as MetaCyc \citep{caspi2020metacyc}.
When such information is not available for the target metabolite, we can select metabolites that are strongly correlated with the target metabolite of interest into the pool. 
However, the initial pool may contain many metabolites, and as shown in Theorem \ref{thm1}, 
using too many auxiliary metabolites could harm the estimation accuracy. 
Therefore, a data-driven procedure to select the best auxiliary metabolites from the initial pool is desired in real applications. 


We develop a cross-validation procedure based on a metric of ``microbial correlation". 
Specifically, let $\{y^{(j)}\}_{j=1}^q$ represent the set of $q$ auxiliary metabolites in the pool with $y^{(0)}$ denoting the target metabolite, 
and let $\{x_k\}_{k=1}^p$ denote all $p$ microbes. 
We first calculate the correlation between the $j$-th metabolite and the $k$-th microbe $r_{jk} = \widehat{\mbox{Cor}}(y^{(j)}, x_k)$
for $j = 0,1, \ldots, q$ and $k = 1, \ldots, p$. Then, we define the ``microbial correlation" $\widehat{\rho}_j$ between the $y^{(0)}$ and $y^{(j)}$ as the correlation between $\{r_{0k}\}_{k=1}^p$ and $\{r_{jk}\}_{k=1}^p$ for each $j=1, \ldots, q$. 
This microbial correlation essentially characterizes the similarities between different columns of the heatmap in   Fig. \ref{Fig1}A. 
Let $\widehat{r}_j = |\widehat{\rho}_j|$ denote the absolute value of each correlation coefficient $\widehat{\rho}_j$ for $j = 1,\ldots, q$. Let $\widehat{p}_j$ denotes the $p$-value corresponding to each $\widehat{\rho}_j$ for $j = 1,\ldots q$. The optimal set of auxiliary metabolites is selected in four steps.
	\begin{enumerate}
		\item Given some threshold $r_0$ (say 0.5) and some significance level $p_0$ (say 0.05), 
only keep the metabolites meeting both conditions in the metabolome pool: $\mathcal{S} = \{j: \widehat{r}_j > {r}_0 \text{ and } \widehat{p}_j <{p}_0\}$ and let $q_0 = |\mathcal{S}|$.
  \item Sort $\{\widehat{r}_j: j \in \mathcal{S}\}$ in non-increasing order: $\widehat{r}_{j_1} \geq \widehat{r}_{j_2} \geq \cdots \geq \widehat{r}_{j_{q_0}}$.  
  \item For each $m = 1, \ldots, q_0$, 
based on $k$-fold cross-validation, evaluate the prediction performance of CoMMiT using $\mathcal{I}_m = \{y^{(j_1)}, \ldots, y^{(j_m)}\}$ as the set of auxiliary metabolites according to the mean squared error, denoted by $\mbox{MSE}_m$. 
	\item Find $m_{\text{opt}} = \mbox{argmin}_m \mbox{MSE}_m$, and obtain the optimal set of informative metabolites $\mathcal{I}_{m_{\text{opt}}} = \{y^{(j_1)},\ldots, y^{(j_{m_{\text{opt}})}}\}$.
	\end{enumerate}

\section{Simulations}\label{sec:simulation}
For each $i=1,\ldots,n$, we generated the CLR-transformed microbiome abundances ${\bf x}_i \in \mathbb{R}^{p}$ from a multivariate normal distribution with mean $\bf 0$ and covariance $\Sigma$, where $\Sigma = (\sigma_{jk})_{jk}$ with $\sigma_{jk} = \rho^{|j-k|}$ for $j,k = 1, \ldots, p$. We considered two auxiliary metabolites. 
Specifically, we generated \[\ve \beta^{(1)} = (\underbrace{\theta_1, \ldots, \theta_1}_{s}, \underbrace{0, \ldots, 0}_{p-2s}, \underbrace{\theta_1, \ldots, \theta_1}_{s} )^\intercal \mbox{ and } \ve \beta^{(2)} = (\underbrace{0, \ldots, 0}_{s}, \underbrace{\theta_1, \ldots, \theta_1}_{2s}, \underbrace{0, \ldots, 0}_{p-3s})^\intercal;\] here, $s$ controls the sparsity, and $\theta_1$ characterizes the signal size. 
Next, for $k=1,2$, we generated the $k$-th auxiliary metabolite abundance according to $y_{i}^{(k)} = {\bf x}_i^\intercal \ve \beta^{(k)} + \varepsilon^{(k)}_{i}$, where $\varepsilon^{(k)}_{i} \stackrel{i.i.d}{\sim} N(0,\sigma_k^2)$ for $i = 1, \ldots, n$. Since ${\ve \beta^{(1)}}^\intercal \ve \beta^{(2)} = 0$, the two auxiliary metabolites were designed to carry distinct information. 
The parameter associated with the target metabolite was set as  \(\ve \beta^{(0)} = \ve \beta^{(1)} - 0.5\ve \beta^{(2)} + \ve \omega,\)
where \[\ve \omega = \big(\underbrace{0, \ldots, 0}_{3s}, \underbrace{\theta_2, \ldots, \theta_2}_{s}, \underbrace{0, \ldots, 0}_{p-4s}\big)^\intercal.\] Hence, a smaller $\theta_2$ indicates a better alignment between $\ve \beta^{(0)}$ and the space spanned by $\ve \beta^{(1)}$ and $\ve \beta^{(2)}$. 
Then, the target metabolite abundance was generated according to $y^{(0)}_i = {\bf x}_i^\intercal \ve \beta^{(0)} + \varepsilon_i^{(0)}$ with $\varepsilon^{(0)}_i \stackrel{i.i.d}{\sim} N(0, \sigma_0^2)$ for $i = 1, \ldots, n$. 

We compared our CoMMiT with several state-of-the-art high-dimensional estimation and inference methods as well as recent transfer-learning methods. 
For estimation, we considered Lasso \citep{tibshirani1996regression}, Ridge \citep{hoerl1970ridge}, Trans-lasso \citep{li2020transfer, tian2022transfer}, and Angle-TL \citep{gu2024robust}.
For inference, we considered low-dimensional projection estimator (LDPE, \citealp{zhangzhang2014}), Ridge-based inference (Ridge, \citealp{buhlmann2013}), Trans-lasso inference \citep{tian2022transfer}. We did not consider Angle-TL for inference because it only focuses on estimation and prediction. 
For implementation, we used the R packages {\tt glmnet} for both Lasso and Ridge estimation, {\tt hdi} for LDPE and Ridge-based inference, {\tt glmtrans} for Trans-lasso estimation and inference, and {\tt multiTL} for Angle-TL.
	
\begin{figure}[ht!]
\centering
\includegraphics[width=16cm,scale=0.5]{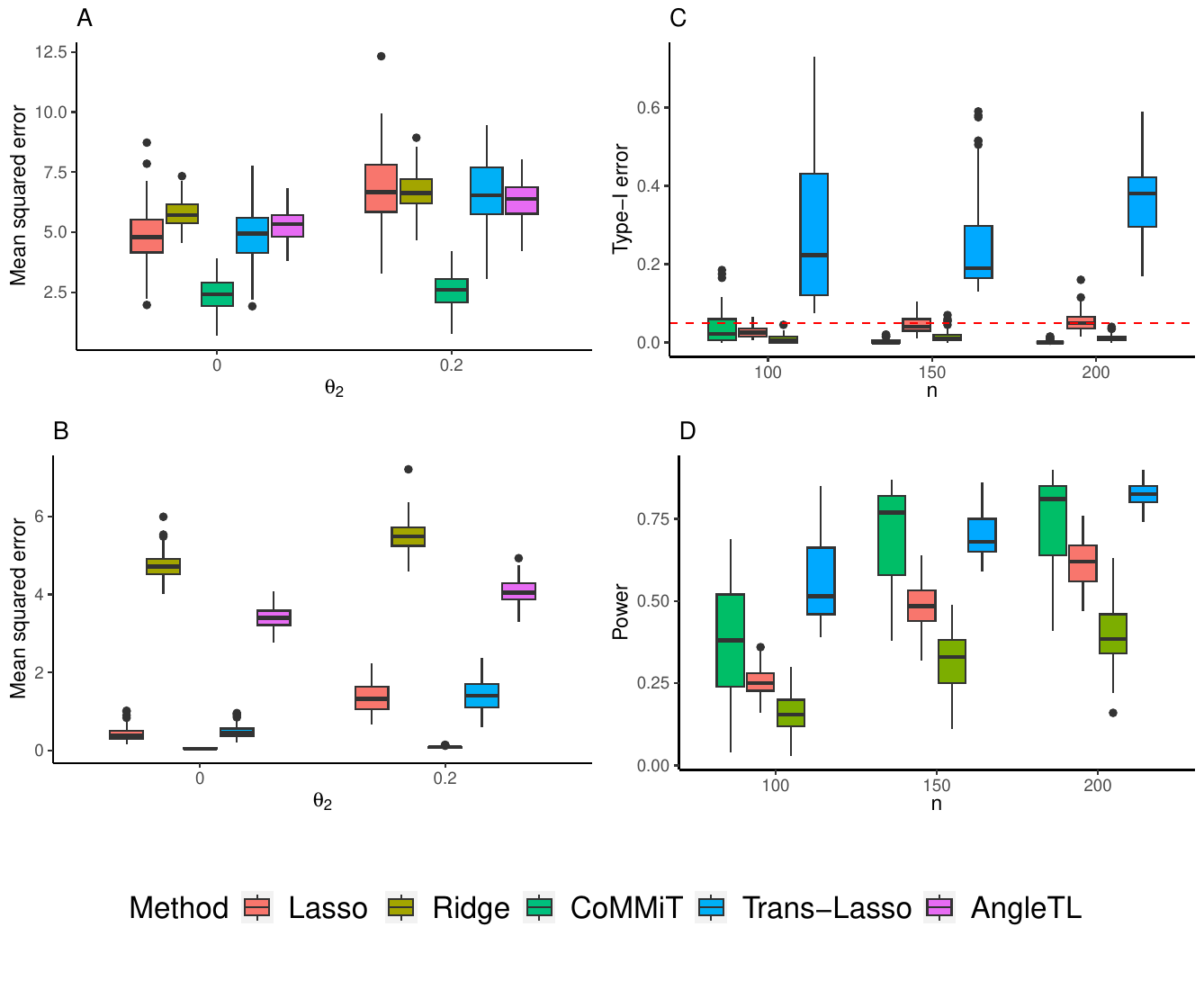}
\caption{Boxplots of the mean squared error (A) and (B) for $n = 100, 200$, type-I error (C), and power (D) for $n = 100, 150, 200$ over 100 replications for simulation data with $\theta_2 = 0, 0.2$, respectively: the proposed CoMMiT estimation has lower MSE, indicating the superior estimation accuracy over other existing methods. All methods except for Trans-Lasso can control the type-I error rate. Among the methods that can control the type-I error rate, CoMMiT has the highest power. }
\label{Fig3}
\end{figure}

We first compared the proposed CoMMiT estimation with the selected methods in terms of the mean-squared error (MSE) for estimating $\ve \beta^{(0)}$. We set $n=100$ and $200$, $p=300$, $\theta_1=0.3$, $\theta_2=0, 0.2$, $s=20$, $\sigma_0=0.2$, $\sigma_1=\sigma_2=0.1$, and $\rho=0.3$. The results are shown in Fig. \ref{Fig3}. It can be seen that CoMMiT has superior estimation accuracy over all existing methods. 
In particular, CoMMiT outperforms the two existing transfer-learning methods (Trans-lasso and Angle-TL) which also leverage information from the auxiliary metabolites. 
This is because, by design, neither of the two auxiliary outcomes is strongly informative for the target outcome, leading to the sub-optimal performance of Trans-lasso and Angle-TL. 
In contrast, CoMMiT can efficiently leverage the joint informativeness of the auxiliary metabolites without requiring individual auxiliary metabolite to be informative. 
In Fig. \ref{Fig3}A, with $n=100$, both Lasso and Ridge show higher or similar MSEs to Trans-Lasso and AngleTL. However, Fig. \ref{Fig3}B reveals that when $n = 200$, Lasso outperforms Trans-Lasso and Angle-TL.
This indicates the possibility of having a negative transfer (i.e., a decline in learning performance due to using source data) if the auxiliary information is not used properly. 
Lasso performs better than Ridge due to the sparsity of the target parameter. 
In terms of running speed, Angle-TL has the longest average runtime, taking over 136 seconds (when $n=200$) compared to CoMMiT ($0.27$ seconds) and Trans-lasso ($1.03$ seconds) over 100 replications. 

We then compared our proposed CoMMiT inference with the selected inference methods in terms of the type-I error rate and power. We set $\theta_1 = 1$ to increase the signal size, because, under the previous setting with $\theta_1=0.3$, all the methods considered here would lead to very low power. Accordingly, we increased the variance of the error terms by setting $\sigma_0=2$ and $\sigma_1=\sigma_2=1$, and considered $\theta_2=0.4$ under simulation $n = 100, 150$, and $200$. 
We evaluated the power of each method by testing the non-zero coefficients of $\ve \beta^{(0)}$ and the type-I error rate by testing the zero coefficients of $\ve \beta^{(0)}$. 
We considered a two-sided significance level $\alpha=0.05$. The results are presented in Figs. \ref{Fig3}C and D. It can be seen from Fig. \ref{Fig3}C that all methods, except for Trans-lasso, can effectively control the type-I error rate. More specifically, we see that CoMMiT is the most conservative when $n = 150$ and 200.
Nonetheless, as seen in Fig. \ref{Fig3}D, CoMMiT still has the highest power among the methods that can control the type-I error rates. 
\section{Real Data}\label{sec:realdata}

We revisited the CARB study to systematically explore the interactions between gut microbes and TUDCA as well as other related bile acids by leveraging information from other bile acids under the LGL diet using CoMMiT. 
As outlined in Section \ref{sec:carb}, among these 29 bile acids, four were secondary bile acids: UDCA, DCA, HDCA, and LCA. These acids, crucial for lipid digestion and intestinal health, are formed in the colon from primary bile acids through bacterial metabolism \citep{ajouz2014secondary}. However, certain types like DCA and LCA can be harmful in excess, potentially leading to a higher risk of colon cancer and other gastrointestinal diseases \citep{ajouz2014secondary, farhana2016bile}.

Given that TUDCA is derived from UDCA, a cytoprotective agent \citep{im2004ursodeoxycholic}, we began by investigating the impact of gut microbes on UDCA. Initially, a univariate regression model was applied to each microbe's abundance in relation to UDCA. As illustrated in Fig. \ref{Fig5}A, $25$ gut microbes exhibited significant marginal associations with UDCA, suggesting their involvement in UDCA production. 
Building on these marginal correlations, we next employed existing high-dimensional inferential methods (LDPE and Ridge inference) to gain biologically more meaningful results. 
LDPE identified $10$ significant microbes (Fig. \ref{Fig5}B), while Ridge found one (Fig. \ref{Fig5}C), indicating LDPE's higher statistical lower over Ridge. 
However, in terms of prediction accuracy measured by mean squared error (MSE) in leave-one-out cross validation (LOOCV), Ridge outperformed Lasso (Ridge: $1.12$ vs. Lasso: $1.21$). This discrepancy between prediction and inference is possibly due to Ridge's conservative nature in high-dimensional settings \citep{buhlmann2013}.
Both LDPE (Lasso) and Ridge will serve as the benchmark in this analysis.

CoMMiT was then utilized to examine UDCA-microbiome interactions, leveraging auxiliary bile acids selected by the proposed method in Section \ref{sec:heuristic}. 
Specifically, with an initial pool of auxiliary metabolites based on ${r}_0 = 0.5$ and $p_0 = 0.01$, the optimal auxiliary metabolite set comprised four bile acids, including two secondary acids ({DCA} and {LCA}), a conjugated acid ({Glycodeoxycholic Acid (GDCA)}), and a modified form of LCA ({3$\alpha$-Hydroxy-12-Ketolithocholic Acid}). 
These acids have varied relationships with UDCA according to the existing literature. 
For example, UDCA and its primary metabolite LCA work together to guard against intestinal inflammation, achieving this through the inhibition of epithelial cell apoptosis and enhancement of barrier integrity in vivo \citep{lajczak2020secondary}; DCA is associated with an increased risk of colon cancer, UDCA is noted for its potential protective properties against it \citep{centuori2014differential}; GDCA accelerates the growth of polycystic human cholangiocytes; However, this proliferative effect is counteracted by UDCA and its derivative TUDCA \citep{munoz2015ursodeoxycholic}.

\begin{figure}[!ht]
	\centering
	    \includegraphics[width=1.0\textwidth,height=0.35\textheight]{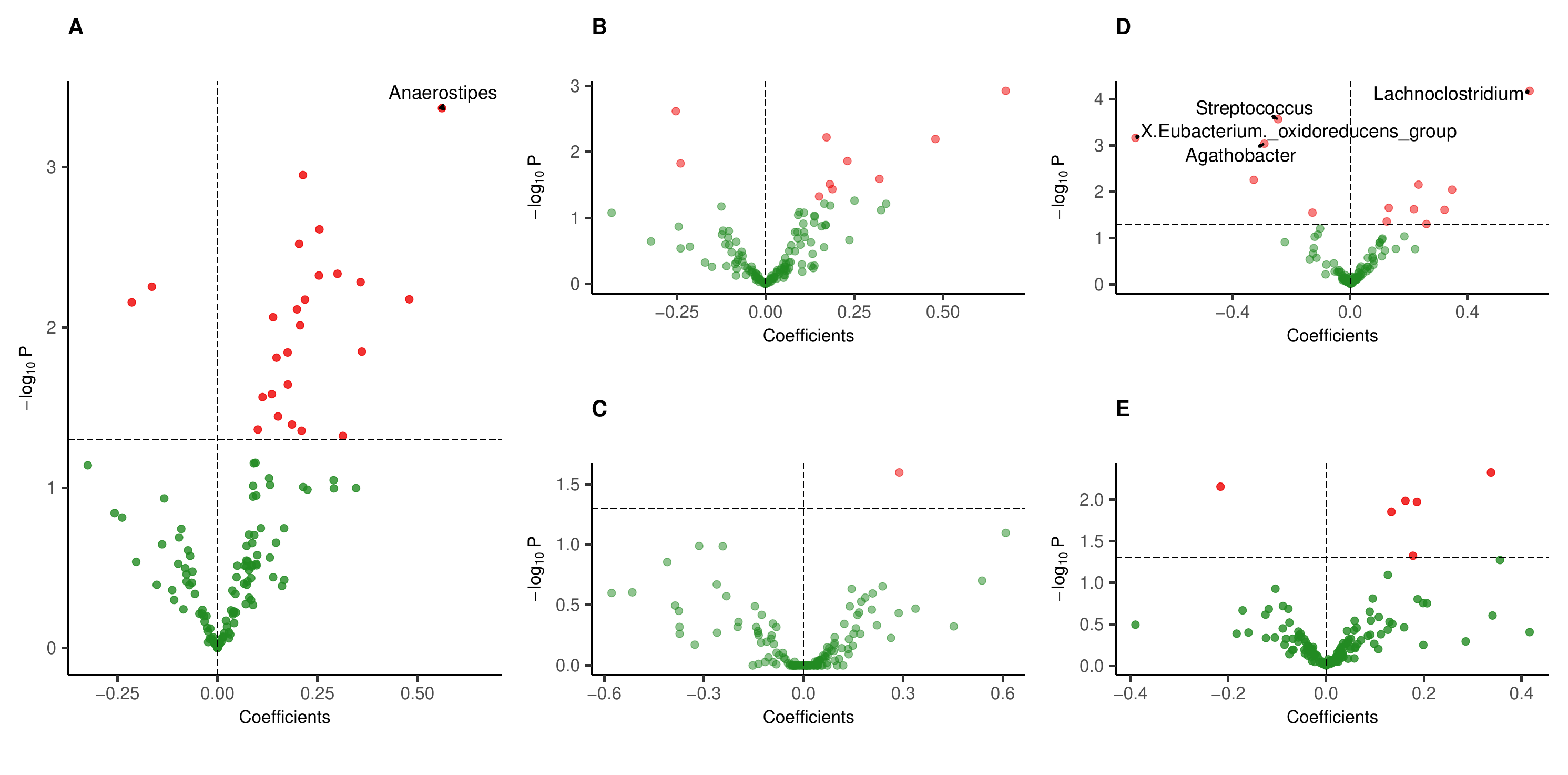}
	\caption{ Volcano plots of the $-\log_{10}(\text{p value})$ (y-axis) and regression coefficients (x-axis) showing the association between microbes and UDCA using univariate linear regression (A), LDPE (B), Ridge-based inference (C), CoMMiT (D), and Trans-lasso (E). 
 The dashed vertical and horizontal lines represent the filtering criteria: fold change at $0$ and cut off at $-\log_{10}(0.05)$, respectively. 
 Colors indicate the significance of individual gut microbes: red dots represent significant gut microbes (with $p$-value $< 0.05$), and green dots represent non-significant gut microbes. To enhance readability, only microbes with $p$-value $< 0.001$ are labeled.}
	\label{Fig4}
\end{figure}

\begin{figure}[H]
	\centering
	    \includegraphics[width=1.0\textwidth,height=0.35\textheight]{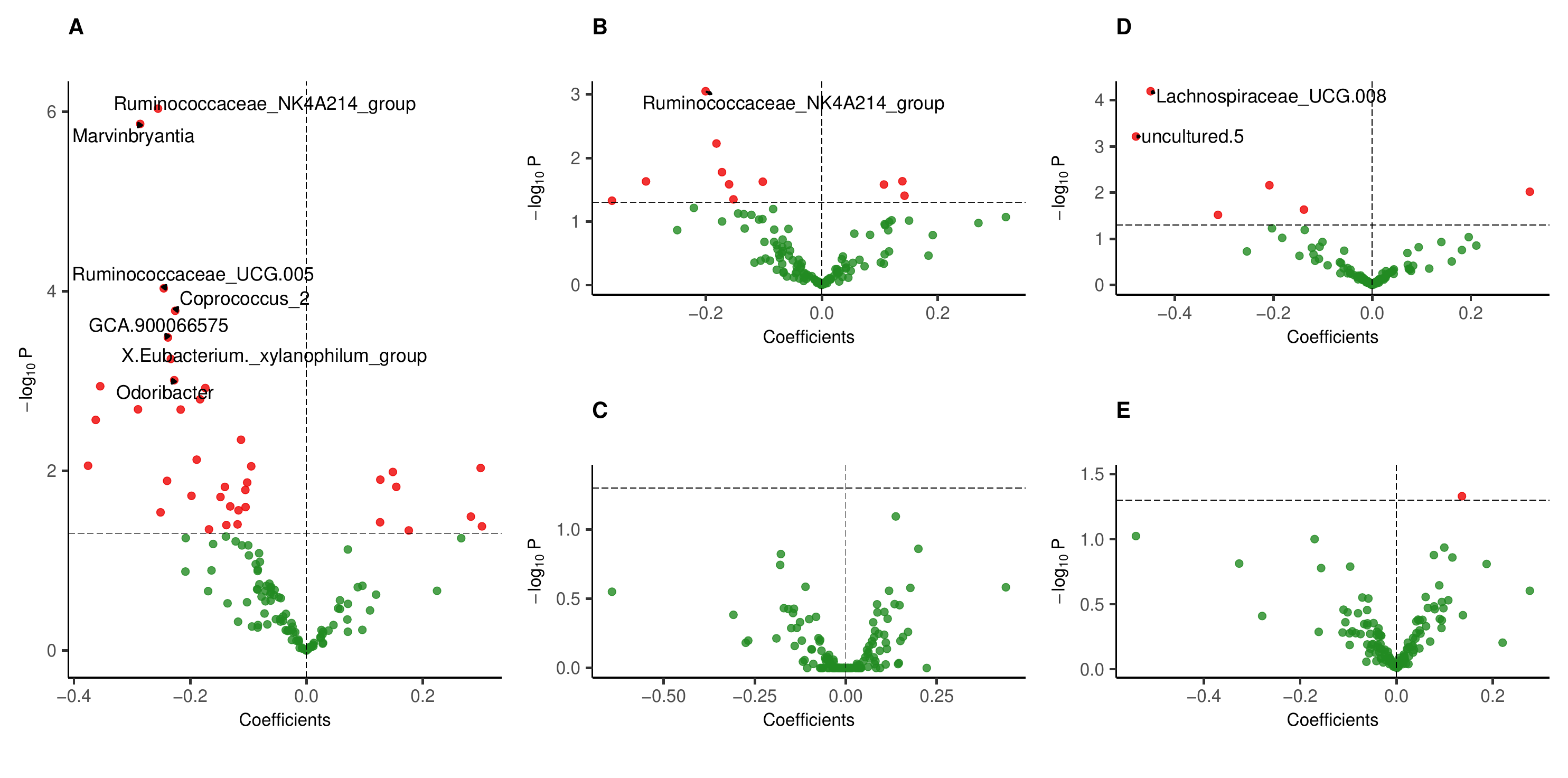}
	\caption{Volcano plots of the $-\log_{10}(\text{ p value})$ (y-axis) and regression coefficients (x-axis) showing the association between microbes and TUDCA using univariate linear regression (A), LDPE (B), Ridge-based inference (C), CoMMiT (D), and Trans-lasso (E). 
 The dashed vertical and horizontal lines represent the filtering criteria: fold change at $0$ and cut off at $-\log(0.05)$, respectively. 
  Colors indicate the significance of individual gut microbes: red dots represent significant gut microbes (with $p < 0.05$), and green dots represent non-significant gut microbes. To enhance readability, only microbes with $p < 0.001$ are labeled.}
	\label{Fig5}
\end{figure}

We compared the prediction accuracy of CoMMiT with Lasso, Ridge, Trans-Lasso, and Angle-TL in terms of the mean-squared error (MSE). 
For fair comparison, Trans-Lasso and Angle-TL were also implemented by using the same four auxiliary metabolites as those used for CoMMiT.
The results show that CoMMiT had the lowest MSE (1.11), indicating superior prediction accuracy over the existing methods. 
While Trans-Lasso (1.14) and Angle-TL (1.15) had slightly lower MSEs compared to Lasso (1.21), they both performed slightly worse compared to Ridge (1.12). 
We then examined the conditional associations between the microbes and UDCA using our proposed CoMMiT inference as well as Trans-Lasso. 
The results are summarized in Figs. \ref{Fig4}D and E, respectively. It can be seen that CoMMiT (Fig. \ref{Fig4}D) identified $13$ statistically significant gut microbes, whereas Trans-Lasso only identified $6$ (Fig. \ref{Fig4}E). 
This result is different from Trans-Lasso's markedly inflated type-I error rate in Fig. \ref{Fig3},
indicating that 
the potential effect of negative transfer might be different for Trans-Lasso in different applications. 

All methods in Fig. \ref{Fig4}, except for Ridge, identified {Streptococcus}. The univariate regression model, LDPE, and CoMMiT all identified {Lachnoclostridium}. The former, {Streptococcus}, is a bacteria closely associated with liver cirrhosis \citep{zhong2021streptococcus} and other types of infections such as throat and skin infections. The latter, {Lachnoclostridium}, is genus of bacteria in the {Lachnospiraceae} family, which is responsible for conversion of primary to secondary bile acids. 
Notably, after FDR adjustment at the level of 0.05, {Lachnoclostridium} remained significant for CoMMiT, whereas it was no longer identified as significant by other methods.
Moreover, after accounting for FDR at the level 0.05, 
we observed that no significant gut microbes were detected as significant using existing methods,
whereas for CoMMiT, three gut microbes survived FDR at $0.05$ including both {Streptococcus} and {Lachnoclostridium}; in fact, {Lachnoclostridium} survived a more stringent FDR control at 0.01.


Building upon our findings regarding the interactions between UDCA and gut microbes, we extended our investigation to explore the interactions between TUDCA and the microbiomes. Following the approach detailed in Section \ref{sec:heuristic}, GHDCA, a glycine-conjugated form of HDCA, was selected as the sole auxiliary outcome. The microbial correlation between TUDCA and GHDCA is notably high at 0.92, marking it as the strongest correlation among all bile acids studied. Intriguingly, the microbial correlation between UDCA and TUDCA is relatively weak, at only $-0.06$. 
This may be because UDCA is made from CDCA in the gut by the microbes and then converted to TUDCA by human enzymes in the liver. Thus, bacteria are not directly involved in the conversion of UDCA to TUDCA, which may explain the weak microbial correlation between UDCA and TUDCA.

In evaluating prediction accuracy through LOOCV, CoMMiT demonstrated superior performance to other existing methods in terms of mean squared error (MSE). Specifically, CoMMiT achieved an MSE of 0.63, which was notably lower than that of Lasso (0.73), Ridge (0.66), Trans-Lasso (0.68), and Angle-TL (0.70). It is important to note that in this scenario, with only one auxiliary outcome, our projection-based distance (as shown in Fig. \ref{Fig2}A) corresponds to the angle-based distance (illustrated in Fig. \ref{Fig2}C). However, despite this similarity, Angle-TL still resulted in a higher MSE than CoMMiT. This divergence in performance may be attributed to the distinct algorithmic approach of Angle-TL, particularly its use of a $\sin \Theta$-type penalty.

In the context of inference, $38$ gut microbes were found to be marginally associated with TUDCA, as shown in Fig. \ref{Fig5}A. The LDPE method identified $9$ of these as conditionally significant (Fig. \ref{Fig5}B), whereas Ridge inference did not identify any  significant associations (Fig. \ref{Fig5}C). CoMMiT identified $6$ significant gut microbes (Fig. \ref{Fig5}D), while Trans-Lasso flagged identified $1$ (Fig. \ref{Fig5}E). The {Ruminococcaceae NK4A214 group}, identified by LDPE, was also detected by CoMMiT. Notably, despite LDPE identifying more microbes than CoMMiT – a contrast to the simulation results in Fig. \ref{Fig3}D where CoMMiT exhibited higher power – only three microbes were commonly identified by both CoMMiT and LDPE: {Ruminococcaceae NK4A214 group}, {uncultured 5}, and {Lachnospiraceae\_UCG.008}. This suggests that LDPE might have reported some false positives.

CoMMiT uniquely identified $3$ gut microbes, each notable for their roles in gut health and interactions with secondary bile acids.
There include {GCA-900066575} and {Fusicatenibacter}, members of the {Lachnospiraceae} family;  {X Eubacterium hallii group}, a subgroup within the genus {Eubacterium}.
Existing research highlights the connections between these microbes and secondary bile acids. 
For instance, 
a recent study on pediatric nonalcoholic fatty liver disease found that reduced levels of {Eubacterium} is correlated with secondary conjugated bile acids such as TUDCA \citep{yu2021disease}.

\section{Discussions}\label{sec:discussions}
This paper introduces CoMMiT, a novel within-cohort transfer-learning framework designed to investigate microbiome-metabolome interactions with limited data and weak biological signals. 
CoMMiT leverages information from auxiliary metabolites within the same study that are collectively informative for the target metabolite. Unlike existing transfer-learning methods, CoMMiT does not require individual auxiliary metabolites to be informative or rely on external datasets which are potentially heterogeneous, thereby minimizing the risk of negative transfer. 
Through extensive simulations and real-world applications, CoMMiT demonstrates significant advantages over state-of-the-art high-dimensional models and existing transfer-learning models. 
We anticipate broad applicability to other datasets with similar dependency structures, especially in cases of significant heterogeneity across studies, where transferring knowledge within a cohort is more justified.

While our focus has been on linear models, driven by targeted metabolomics data, extending CoMMiT to more complex models offers valuable opportunities. For instance, in untargeted metabolomics, where excessive zeros (missing data) are common, a mixture model could differentiate between zero and non-zero values. For a given metabolite $y$, with $\delta=1$ indicating a non-zero value and $\delta=0$ for zeros, the model can be expressed as follows:
\begin{align*}
& \mathbb{P}(\delta=0)=1-p+p \Phi \left( \frac{T-\mu}{\sigma} \right) \\
& \mathbb{P}(z,\delta_i=1)= p \frac{1}{\sqrt{2\pi \sigma^2}} \exp \left ( - \frac{1}{2\sigma^2} (\log(y)-\mu)^2  \right), \,\, \textnormal{for } \log(y) \geq T,
\end{align*}
where $\Phi(\cdot)$ is the normal CDF, $p$ indicates the presence probability of the metabolite, $T$ represents the detection limit, and $\mu$ and $\sigma$ are the mean and standard deviation, respectively, of the metabolite's value when present. We link $\mu$ to the microbes as $\mu = {\bf x}^\intercal \ve \beta$. Extending CoMMiT to this mixture model framework is a promising direction for future research.

The ``microbial correlation" method in Section \ref{sec:heuristic}, while effective for selecting informative metabolites, is grounded in marginal correlations rather than conditional associations. A formal evaluation of this microbial correlation is thus of interest. In the context of the linear model \eqref{eq:0}, we define the Information Score (IS) between $y^{(0)}$ and $y^{(j)}$ as the cosine of the angle between $\ve \beta^{(0)}$ and $\ve \beta^{(j)}$, expressed as
\[
\text{IS}_j = \left|\cos\left( \frac{\ve \beta^{(0) \intercal} \ve \beta^{(j)}}{ \|\ve \beta^{(0)}\| \cdot \|\ve \beta^{(j)}\| }\right)\right|.  
\]
A higher $\text{IS}_j$ value suggests that $y^{(j)}$ is more informative for $y^{(0)}$. While similar approaches are used in genetic correlations \citep{guo2019optimal}, genome-wide association studies typically have larger sample sizes than multi-omics microbiome studies. Exploring the efficacy of the information score in identifying informative metabolites, in comparison to the ``microbial correlation" is an intriguing avenue for research.


\bigskip
\begin{center}
{\large\bf SUPPLEMENTARY MATERIAL}

The supplementary materials contain proof of the theoretical results.
\end{center}

\bibliographystyle{chicago}
\bibliography{Bibliography-MM-MC, ref-r03-1, ref-r03-2, ref-r03-3, reference_wang_ye, reference_NIH}

\end{document}